\documentclass[conference]{IEEEtran}
\IEEEoverridecommandlockouts
\usepackage{cite}
\usepackage{amsmath,amssymb,amsfonts}
\usepackage{algorithmic}
\usepackage{graphicx}
\usepackage{textcomp}
\usepackage{xcolor}
\usepackage{tablefootnote}
\usepackage{subfigure}
\def\BibTeX{{\rm B\kern-.05em{\sc i\kern-.025em b}\kern-.08em
    T\kern-.1667em\lower.7ex\hbox{E}\kern-.125emX}}
\begin{document}

\title{An investigation of licensing of datasets for machine learning based on the GQM model\\
}

\author{\IEEEauthorblockN{Junyu Chen}
\IEEEauthorblockA{\textit{Nagoya University}\\
Nagoya, Japan \\
chen.junyu.v3@s.mail.nagoya-u.ac.jp}
\and
\IEEEauthorblockN{Norihiro Yoshida}
\IEEEauthorblockA{\textit{Ritsumeikan University}\\
Kusatsu, Japan \\
norihiro@fc.ritsumei.ac.jp}
\and
\IEEEauthorblockN{Hiroaki Takada}
\IEEEauthorblockA{\textit{Nagoya University}\\
Nagoya, Japan \\
hiro@ertl.jp}
}

\maketitle

\begin{abstract}
Dataset licensing is currently an issue in the development of machine learning systems. And in the development of machine learning systems, the most widely used are publicly available datasets. However, since the images in the publicly available dataset are mainly obtained from the Internet, some images are not commercially available. Furthermore, developers of machine learning systems do not often care about the license of the dataset when training machine learning models with it. In summary, the licensing of datasets for machine learning systems is in a state of incompleteness in all aspects at this stage.

Our investigation of two collection datasets revealed that most of the current datasets lacked licenses, and the lack of licenses made it impossible to determine the commercial availability of the datasets. Therefore, we decided to take a more scientific and systematic approach to investigate the licensing of datasets and the licensing of machine learning systems that use the dataset to make it easier and more compliant for future developers of machine learning systems.
  
This paper applies the GQM method, which is a framework originally used for systematic metrics and analysis in the field of software engineering, to investigate the licensing of datasets and the licensing of machine learning systems. We designed two questions, ``Are there licensing issues in the machine learning dataset?" and ``Is it easy to use machine learning datasets following the licensing?" as the final goal of this investigation, and designed 7 related questions and 12 quantifiable metrics to investigate and answer. To solve the final goal, we investigated 311 repositories related to machine learning systems on GitHub, with a total of 42 datasets used by these systems. By investigating the licensing of these repositories and datasets, we found that in the present environment of lack of well-managed dataset licensing, developers of machine learning systems are exposed to potential licensing violations if they want to use publicly available datasets.
\end{abstract}


\section{Introduction}
In this section, we will introduce the background of the current research on licensing of machine learning datasets. Next, we will introduce our target based on this background and our contribution.
\subsection{Background}
The development of machine learning systems in the last decade has been exponential. And in machine learning systems, the essential thing is the models trained based on datasets. Machine learning datasets are collections of data used to train and evaluate machine learning models. So as machine learning development grows and becomes huge in scalability, the use of datasets becomes increasingly popular. Therefore, the collection and development of datasets are particularly needed. However, as the field of machine learning has shifted to approaches with more significant data requirements over the past decade, the skilled and methodical annotation applied in early dataset collection practices was considered slow and consuming. Today's data collection tends to shift to unconstrained data, which has led to more and more data collection from the web \cite{halevy2009unreasonable}. These datasets can come from a variety of sources, such as government agencies, research institutions, and private companies. 

 Since traditional data collection and annotation are considered to be slow and costly, most developers of datasets use script-like tools to collect data directly from the Internet. Most of this data licensing is not taken seriously by the developers. In order to use these datasets, it is important to understand the terms and conditions of the license under which they are made available. The neglect of licensing of the collected data will result in unknown commercial exploitability of the dataset. In fact, in our investigation, most of the datasets were found to be of unknown commercial use.

Two laws generally govern datasets and data contained in datasets: \emph{copyright law} and \emph{Contract law} \cite{cani}. \emph{Copyright law} prohibits copying or modifying any part of a work without the express permission of the copyright holder. Copyrighted data may not be used or distributed commercially without the express permission of the copyright holder. However, many publicly available datasets are known to contain copyrighted data. Using these data to build commercial AI software may lead to copyright infringement. \emph{Contract law} allows the copyright holder to grant a license outlining the rights of others and their obligations. If rights not granted by the license are exercised over the data, or if obligations are not fulfilled, there may be a contract violation. While the actual protection provisions provided by these laws vary by country/region, these laws typically provide some similar protections. In addition, with the development of artificial intelligence, several privacy and fairness-related laws have also governed the use of datasets. It's essential to understand the license terms and conditions of the dataset you are using so as to ensure you are using the data in accordance with the wishes of the creators and not violating any legal or ethical guidelines.

Moreover, as for known licensing, several different types of licenses can be used for machine learning datasets, each with its own set of terms and conditions. Some standard licenses include the Open Data Commons Public Domain Dedication and License (PDDL), the Creative Commons Attribution License (CC BY), and the GNU General Public License (GPL). We will introduce them in Section 2.3 in detail.

In this paper, we will divide the dataset licensing issues that cause licensing violations in machine learning systems into three categories:
\begin{itemize}
\item{Since the images in the dataset were mainly obtained from the Internet, some images may or may not be commercial, e.g., if the license of a dataset is CC BY-NC, it is non-commercial.}
\item{Datasets may be withdrawn for legal or other reasons, e.g., MS-Celeb-1M \cite{guo2016ms} was withdrawn due to a dispute over portrait rights \cite{MSdataset}.}
\item{A dataset may be withdrawn from the public domain because the original purpose was achieved, e.g., MegaFace \cite{kemelmacher2016megaface} was withdrawn when the corresponding competition was over.}
\end{itemize}

 In the first issue described above, the use of a non-commercial dataset for commercial purposes is considered to result in a license violation. In the last two issues, if the machine learning system continues to use the dataset in the withdrawn condition, it will also be considered as a license violation.
 
\subsection{Targets \& Contributions}
Our primary target in this paper is to investigate dataset licenses and help to develop a system to assist developers with license management.
On the way to reaching our ultimate target, we can also accomplish two objectives at the same time:
\begin{itemize}
\item{Clarify the license of the dataset so that it is easy to know if the machine learning system uses only commercially possible datasets.}
\item{Ensure the traceability of the dataset, i.e., clarify the origin of the data in the dataset to facilitate future development and modification by the dataset developer.}
\end{itemize}

After accomplishing the above targets, this paper has the following contributions. 
\begin{itemize}
    \item {We have initially investigated and organized the licensing of datasets that are frequently used, or used by systems with a high frequency of use, on GitHub.}
    \item {In the collation of the dataset, we simultaneously collated the data sources of the dataset and analyzed the conflicting licensing of the dataset and the data it contains.}
    \item {We investigated the licensing compatibility between the dataset and the system while analyzing and explaining the two ultimate goals of the GQM model for the dataset investigations.}
    \item {At Last, we appeal to the community of machine learning systems and datasets through recommendations to provide a complete license-based system of the dataset used in the absence of a well-defined law.}
\end{itemize}

\section{Related Work}
In this section, we will introduce related studies on licensing of machine learning datasets. These related studies will include software engineering for machine learning, licensing of general software systems, and licensing of machine learning datasets.
\subsection{Software Engineering for Machine Learning}
With the recent development of machine learning, software that uses machine learning is rapidly penetrating society. At the same time, however, traditional software engineering has become entirely ineffective in the face of systems that incorporate machine learning systems. Methodologies for the development, testing, and operation of machine learning software have yet to be established. Engineers on the development front are still getting by on a trial-and-error basis. Therefore, the field of the intersection of machine learning and software engineering has become a research direction that has received attention in recent years. Within this cross-field, there are generally two categories: Software Engineering for Machine Learning (SE for ML) and Machine Learning for Software Engineering (ML for SE).

 Machine Learning for Software Engineering, typically uses machine learning methods to intervene in various development processes or research in software engineering. It is more common to use machine learning methods to test software systems \cite{pei2017deepxplore, zhang2020machine}.

 More relevant to this paper is Software Engineering for Machine Learning, which uses a software engineering approach to make the development of machine learning systems more standardized and measurable. Software engineering for machine learning is the process of designing, developing, and maintaining software systems that support the creation and deployment of machine learning models. This field is becoming increasingly important as machine learning is being used in more and more industries, from healthcare to finance to transportation.  One of the key challenges in software engineering for machine learning is managing the complexity of the models and the data that they operate on. Machine learning models can be very complex, with millions of parameters and a large number of input variables. This complexity can make it difficult to understand how the model is making its predictions, and it can also make it difficult to debug and troubleshoot issues with the model. Machine learning systems are typically found difficult to deliver machine learning system solutions to production environments because of uncertainty \cite{breiman2001random}. It is not easy to know where to start in a machine learning application and whether what you are doing is safe and correct. Thus, in a related thought, recent work has also discussed the impact of using ML-based system applications on ISO security issues \cite{salay2017analysis}. At the same time, multiple efforts have been made in the industry and academics to automate this process by building frameworks and environments that support ML workflows and their experimentation \cite{ amershi2019software, khomh2018software, lo2021systematic}.

Another essential tool for software engineers for machine learning is version control. This allows developers to track changes to the code over time and to collaborate with other developers on the same codebase. This can be especially important when working on large machine learning projects, as multiple developers may need to work on the same codebase at the same time.

\begin{figure*}
    \centering
    \includegraphics[width=\textwidth]{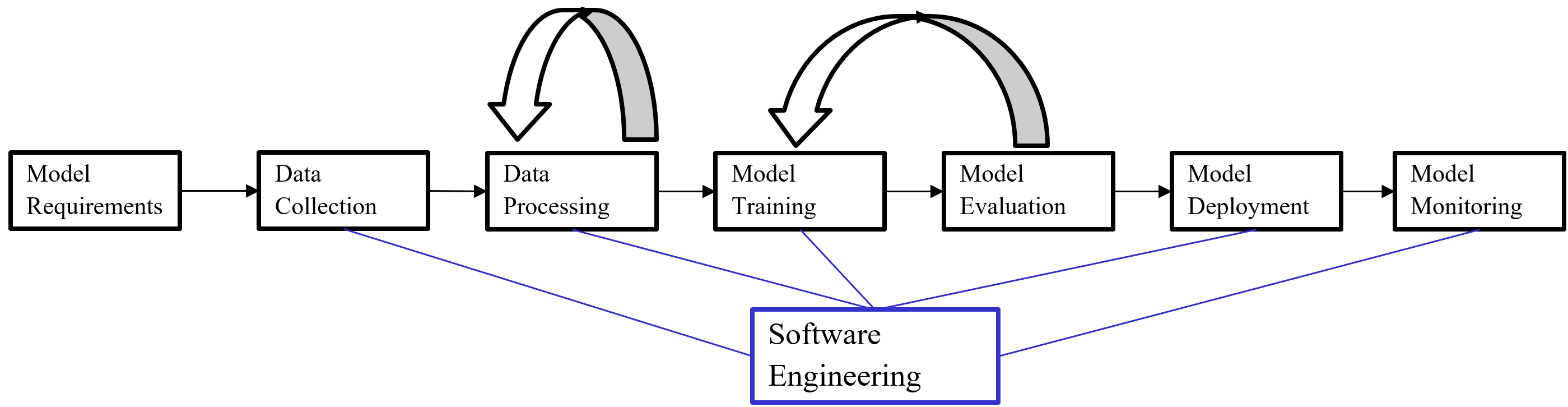}
    \caption{Workflow for machine learning system application development}
    \label{fig:mlworkflow}
\end{figure*}

Moreover, Software Engineering for Machine Learning not only plays a role in the development of machine learning system applications alone but also plays a critical role in data \cite{polyzotis2017data, breueldiscipline}, training \cite{pmlr-v80-kearns18a}, coding \cite{continuous}(see Figure \ref{fig:mlworkflow}). As we can see from the workflow of machine learning system application development, software engineering can influence most of the work steps.

With the development of Software Engineering for Machine Learning, data scientists are often present in the teams of software engineering responsible for the application of machine learning systems \cite{DataScientists}. A challenge in software engineering for machine learning is dealing with the large amounts of data that are required to train and evaluate machine learning models. This data can come from a variety of sources, such as sensors, cameras, and other types of data collection devices. It can also come in a variety of formats, such as images, videos, and audio recordings. Managing this data can be a time-consuming and difficult task, and it can also require specialized software and hardware to store and process the data. This shows that data is essential in the field of Software Engineering for Machine Learning. The investigation mainly done in this paper should also be considered as a part of the data in Software Engineering for Machine Learning.

\subsection{Software Licensing}
Software licensing is the legal agreement that outlines the terms and conditions under which software can be used, distributed, and modified. Software licenses are an important aspect of the software development industry, as they help to protect the rights of the software developer and ensure that the software is used in accordance with the developer's wishes.

Copyright protection of software means that software can only be used with the permission of the copyright owner \cite{kaminski2007open}. Because software can be easily copied, but it is very difficult and expensive to create it \cite{horne2000open}. Currently, there are two primary forms of copyright protection strategies: private and non-private copyright strategies. 

A private copyright policy is a policy that excludes some or all potential users of technology. In a private software development model, code is first protected by copyright and then distributed under a license agreement that grants special rights to users. Therefore, proprietary licenses are the most common type of software license. These licenses are often used for commercial software, and they give the software developer exclusive rights to the software. Under a proprietary license, the software can only be used in accordance with the terms and conditions outlined in the license agreement. This often means that the software can only be used by a single user or organization and that the software cannot be modified or distributed without the developer's permission.

Non-private copyright strategies mainly include placing software in the public domain or licensing it as open source \cite{rosen2005open}. Placing software in the public domain means that the copyright protection of the software is completely waived, and anyone can use and modify it without any compensation, even removing the author's name as their own work. Richard Stallman, a programmer at the Massachusetts Institute of Technology, developed a new method of software distribution, the GNU Public License \cite{stallman2002free}. Stallman's revolutionary ideas about free software subsequently evolved into the current open-source software movement, where the primary purpose of free/open-source software is to maximize openness as well as reduce barriers to the use of software to spread innovation. Open source licenses share software code by assigning rights to intellectual property, which is freely accessible to the user and developer communities to facilitate collaboration and beneficial exchanges between participants with different motivations.

Moreover, freeware licenses are a type of license that allows users to use the software for free, but the software developer retains all rights to the software. Under a freeware license, the software can be used and distributed, but it cannot be modified or sold without the developer's permission. Freeware licenses are often used for software that is intended to be used for personal or non-commercial use.

It is undeniable that with the development of machine learning, more and more private commercial machine learning systems have been put into use and become profitable. However, many machine learning systems are available for public use in open-source software sites, such as GitHub\footnote{https://github.com/}. Of these systems publicly available on the website, some of them are declared to have ``appropriate" open-source software licenses. However, several systems exist that do not have declared software licenses. 

In our investigation, we need to understand the licensing of machine learning systems and consider the possibility of compatibility of the system with the dataset by comparing the commercial availability of the license.

\subsection{Dataset Licensing}
Similar to software, several different types of licenses can be used for machine learning datasets, each with its own set of terms and conditions. Some standard licenses include the Open Data Commons Public Domain Dedication and License (PDDL), the Creative Commons Attribution License (CC BY), and the GNU General Public License (GPL).
\begin{itemize}
    \item {The PDDL is a public domain dedication that allows anyone to use the data for any purpose, without any restrictions or attribution requirements. This license is often used for datasets that are created by government agencies or other public entities, as it allows for maximum accessibility and reuse.}
    \item{The CC BY license is a more restrictive license that requires users to give attribution to the original creator of the dataset when using it. This license is often used for datasets that are created by researchers or other private individuals, as it allows them to retain control over how their data is used.}
    \item{The GPL is a license that is commonly used for software and other forms of open-source code. It allows users to access and use the data, but also requires them to make any modifications they make to the dataset available to the public under the same license. This can be a useful license for datasets that are intended to be used as part of a larger open-source project.}
\end{itemize}

However, relevant research and specifications for licensing of machine learning datasets are exceedingly scarce. Moreover, no previous studies have focused on the licensing violations that occur when building commercial AI software using publicly available datasets \cite{cani}. However, various issues in the dataset can affect the licensing of the dataset. Among these issues, the fairness of the dataset \cite{li2021estimating, DataDebuggingFairness}, the ethical considerations \cite{hanley2020ethical}, and the treatment of personal privacy \cite{yang2022study} are quite explicit in terms of their impact on the relevant laws and licenses. The concepts intersect between these categories, but all have an impact on the statement of the license of the dataset.

\textbf{The fairness of the dataset} is that the machine learning model may be affected by the dataset annotation, which may lead to a severe impact. The success of machine learning is partially attributed to data-driven approaches with large amounts of data, and this end-to-end paradigm makes machine learning models vulnerable to biases in the models themselves and biases in the data. 

\textbf{The ethical considerations of the dataset} refers to the ethical challenges that may arise in the creation and development of datasets for existing societies. This is due to the fact that various processes are prioritized in the creation of a dataset, while other factors, such as data management, and offensive labels, are likely to take a lower priority.

\textbf{The treatment of personal privacy:} For many datasets, the creator of the dataset has collected a large number of images from the Internet that may contain personal information that may not have been permitted by the individual.

Many studies and tools exist to reduce the harm caused by problems with datasets \cite{Foidldatasmell}. For example, to mitigate privacy issues with publicly available image datasets, developers of datasets try to obfuscate private information before releasing the dataset \cite{frome2009large}. Meanwhile, researchers have also been asked to review the ethicality of the dataset \cite{paullada2021data}.

The emphasis on the dataset problem in academia and industry has brought the creation and development of datasets under stricter control, but an established licensing system to regulate datasets is still not in place, as is the case with open-source software.

As mentioned in section 2.2, given the important impact of open-source licenses on open-source development, the field of software engineering has studied its relevant aspects in depth. Therefore, the evaluation of open-source software licenses has also been well-researched \cite{zhang2010automatic, coughlan2020standardizing}. However, the complexity of licensing datasets is different from that of licensing open-source software. Furthermore, the license evaluation of datasets is difficult because the license specification of datasets is not complete. However, in the absence of a dataset licensing specification, Benjamin proposed the Montreal Data License (MDL) for normalizing the licensing of datasets \cite{towardsmon}. Meanwhile, according to Barclay's study, it is becoming increasingly difficult to track data usage and gain a clear understanding of the sources and contributions of data in the system due to the increased complexity of the data ecosystem \cite{2019towardstraceability}. And Hutchinson proposed a framework for operationalizing transparency and accountability based on ethical practices for the development \cite{2021towardsaccountability}.

\section{Preliminary Investigations}
In this section, we will introduce the details and results of two preliminary investigations that were conducted before the main investigation that will be described in this paper. 

Inspired by complaints from researchers in the commercial machine learning community about the licensing of datasets, we decided to conduct a survey of the licensing of publicly available image datasets before deciding to conduct a detailed and complete survey. Therefore, we first conducted the first investigation, i.e., Investigation A. During the discussion of Investigation A, we discovered that we wanted to decide for ourselves the benchmark for finding datasets, i.e., by collating datasets used by machine learning systems on GitHub as objects. 

\subsection{Investigation A}
Prior to Investigation A, we were aware that the vast majority of datasets that could be problematic were visual datasets, i.e., images or videos. The number of image datasets is much larger than the number of video datasets, so we decided to select the investigation targets from the famous image datasets. First, we searched for collections or web pages that collected well-known datasets\footnote{https://ainow.ai/2020/03/02/183280 [Last visited: 
 2023/1/09]}\footnote{https://github.com/awesomedata/awesome-public-datasets\#imageprocessing [Last visited: 2023/1/09]}. Subsequently, we selected 40 datasets from pretty aforementioned websites as survey subjects, such as CelebA\footnote{ http://mmlab.ie.cuhk.edu.hk/projects/CelebA.html[Last visited: 2022/12/08]}, LabelMe\footnote{http://labelme.csail.mit.edu/Release3.0/browserTools/php/dataset.php[Last visited: 2022/12/08]}, MNIST\footnote{http://yann.lecun.com/exdb/mnist[Last visited: 2022/12/08]}, and so on.

We identified and investigated six attributes for each investigation subject, i.e., the datasets. These attributes were:
\begin{itemize}
\item{Name of the dataset}
\item{The place where the dataset can be obtained}
\item{The type of dataset license}
\item{the Place where the dataset license was obtained}
\item{Whether it is still available (not withdrawn)}
\item{Can be used commercially}
\end{itemize}
In identifying and investigating the licenses for these datasets, we found that they fall into five categories: CC series licenses\footnote{https://creativecommons.org/licenses}; ``citation needed" statements (informal licenses); ``not for commercial use" statements (informal licenses); Flickr Terms of Use\footnote{ https://www.flickr.com/help/terms[Last visited: 2022/12/09]}, which indicates that the data source of the dataset is Flickr, an additional classification since images from Flickr may have different licenses; and a large number of ``unknown" licenses.

\begin{figure*}
    \centering
    \includegraphics[width=\textwidth]{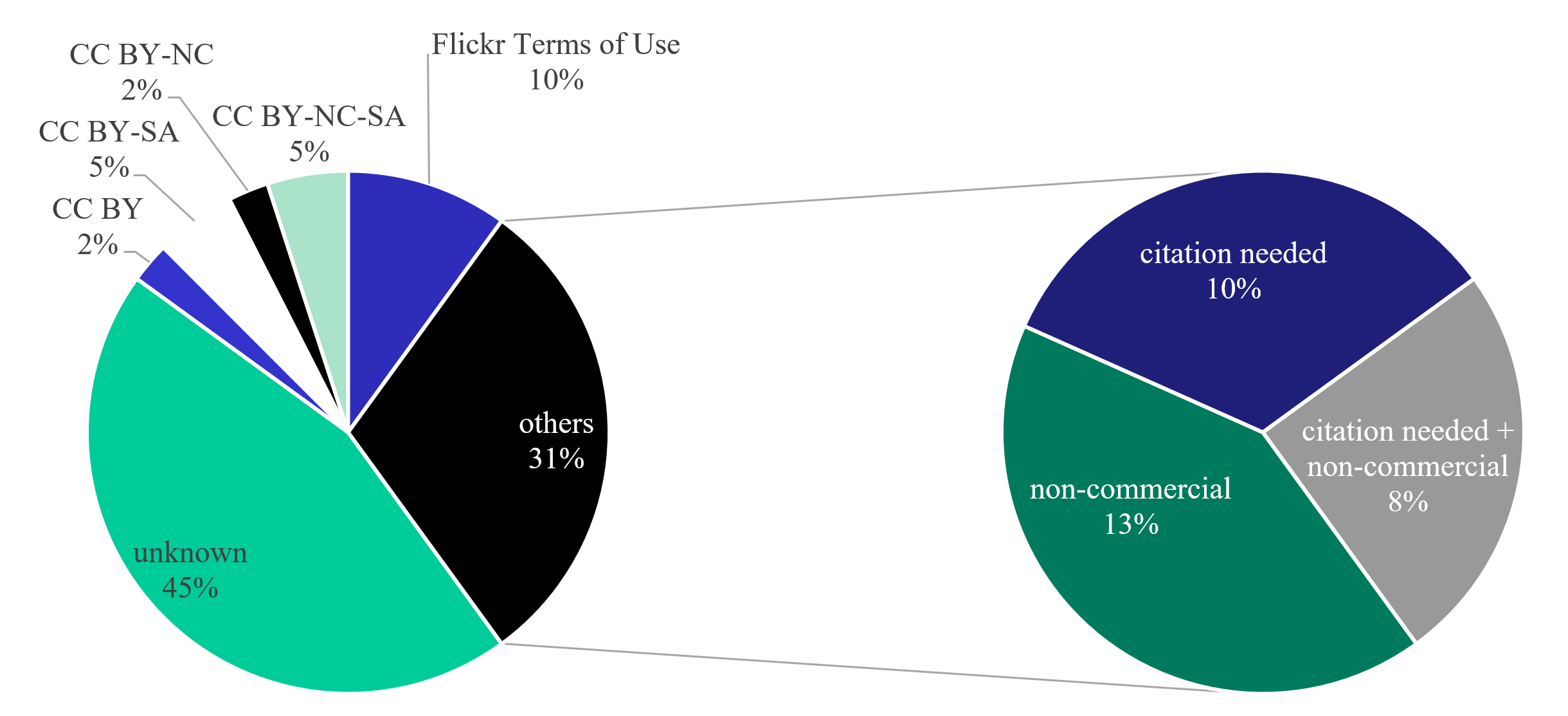}
    \caption{The result of the license of the datasets of Investigation A}
    \label{fig:collectionA}
\end{figure*}

The results we obtained are shown in Figure \ref{fig:collectionA}.

From the pie chart, we can easily see that the licensing of many datasets is unknown, which indicates that many developers of datasets do not even consider the licensing of datasets when publishing data. At the same time, a significant proportion of datasets are declared as ``citation needed" and ``non-commercial". The developers of this part of the public dataset should only declare rather than specifically give a license to the dataset.

In the meantime, one piece of information that fails to appear on top of the pie chart is that the original dataset acquisition URL is no longer accessible for three of the 40 datasets, i.e., it can be assumed that these three datasets were withdrawn.These three datasets are MegaFace\footnote{ http://megaface.cs.washington.edu[Last visted:2022/12/06]}, Food101\footnote{https://vision.ee.ethz.ch/datasets\_extra/food-101[Last visited:2022/12/06]}, and Tiny Images Dataset\footnote{ http://horatio.cs.nyu.edu/mit/tiny/data/index.html[Last visited:2022/12/06]}.

In Investigation A, we treat the commercial availability of Flickr Terms of Use as unavailable; at the same time, we treat the commercial availability of citation needed as unknown. As shown in Table 3.1, in 40 datasets, it is not clear that these datasets are commercially available since the licensing of most of them is unknown. In the remaining 21 datasets, the vast majority of the datasets are explicitly commercially unavailable.
\begin{table*}[h]
\caption{Commercial availability of datasets in Investigation A}
    \centering
    \begin{tabular}{c|c}
    \hline
     Number of commercially available datasets & 3 \\
     \hline
    Number of commercially unavailable datasets &  22\\
     \hline
     Unknown & 19\\
     \hline
    \end{tabular}

\label{tab:collectionA}
\end{table*}

\subsection{Investigation B}
Based on Investigation A, we found that the sources of our targets were collected from the Internet and were insufficiently substantiated. Therefore, in Investigation B, we decided to investigate the repository about machine learning systems on GitHub. The datasets used by these machine learning systems are collated, and then these datasets are used as investigation subjects.

First, we identified five topics related to image machine learning systems: \emph{object-detection}, \emph{semantic-segmentation}, \emph{image-classification}, \emph{recognition}, and \emph{image-annotation}. Afterward, we used GitHub's advanced search feature to investigate the top 30 repositories according to the order of the most stars in each topic as the scope. In the top 30 repositories of each topic, we remove the repositories that are not relevant to the machine learning system and identify only the datasets used in the relevant repositories.

Table 3.2 represents the datasets used by the repositories of machine learning systems under each topic. The number of each item in the table indicates how many repositories under that topic used this dataset. As shown in Table 3.2, among these repositories, the most used datasets are ImageNet, COCO, and Pascal VOC. At the same time, it can be found that not all the repositories under each topic are machine learning systems or declared what dataset was used. Also, several datasets are being used simultaneously by a single machine learning system.

\begin{table*}[h]
\caption{Datasets used by machine learning systems in each topic}
\centering
\begin{tabular}{l|l|l|l|l|l}
\hline
\textbf{topic} & \textbf{\begin{tabular}[c]{@{}l@{}}object\\ detection\end{tabular}} & \textbf{\begin{tabular}[c]{@{}l@{}}semantic\\ segmen-\\ tation\end{tabular}} & \textbf{\begin{tabular}[c]{@{}l@{}}image\\ classi-\\ fication\end{tabular}} & \textbf{\begin{tabular}[c]{@{}l@{}}image\\ annotation\end{tabular}} & \textbf{\begin{tabular}[c]{@{}l@{}}recog-\\ nition\end{tabular}} \\ \hline
ImageNet\tablefootnote{https://image-net.org[Last visited:2022/12/07]}       & 3                                                                   & 1                                                                            & 8                                                                           & 0                                                                   & 2                                                                \\ \hline
COCO\tablefootnote{https://cocodataset.org[Last visited:2022/12/05]}           & 7                                                                   & 3                                                                            & 4                                                                           & 3                                                                   & 0                                                                \\ \hline
Pascal VOC\tablefootnote{http://host.robots.ox.ac.uk/pascal/VOC[Last visited:2022/12/07]}      & 1                                                                   & 7                                                                            & 1                                                                           & 5                                                                   & 1                                                                \\ \hline
Others         & 19                                                                  & 9                                                                            & 17                                                                          & 5                                                                   & 16                                                               \\ \hline
\end{tabular}

    \label{tab:Brepository}
\end{table*}

Through an investigation of a total of 150 repositories, from which we collated, a total of 34 datasets are in use. Moreover, in Investigation B, we find the license CC0 and MIT that Investigation A does not have.

\begin{figure*}
    \centering
    \includegraphics[width=\textwidth]{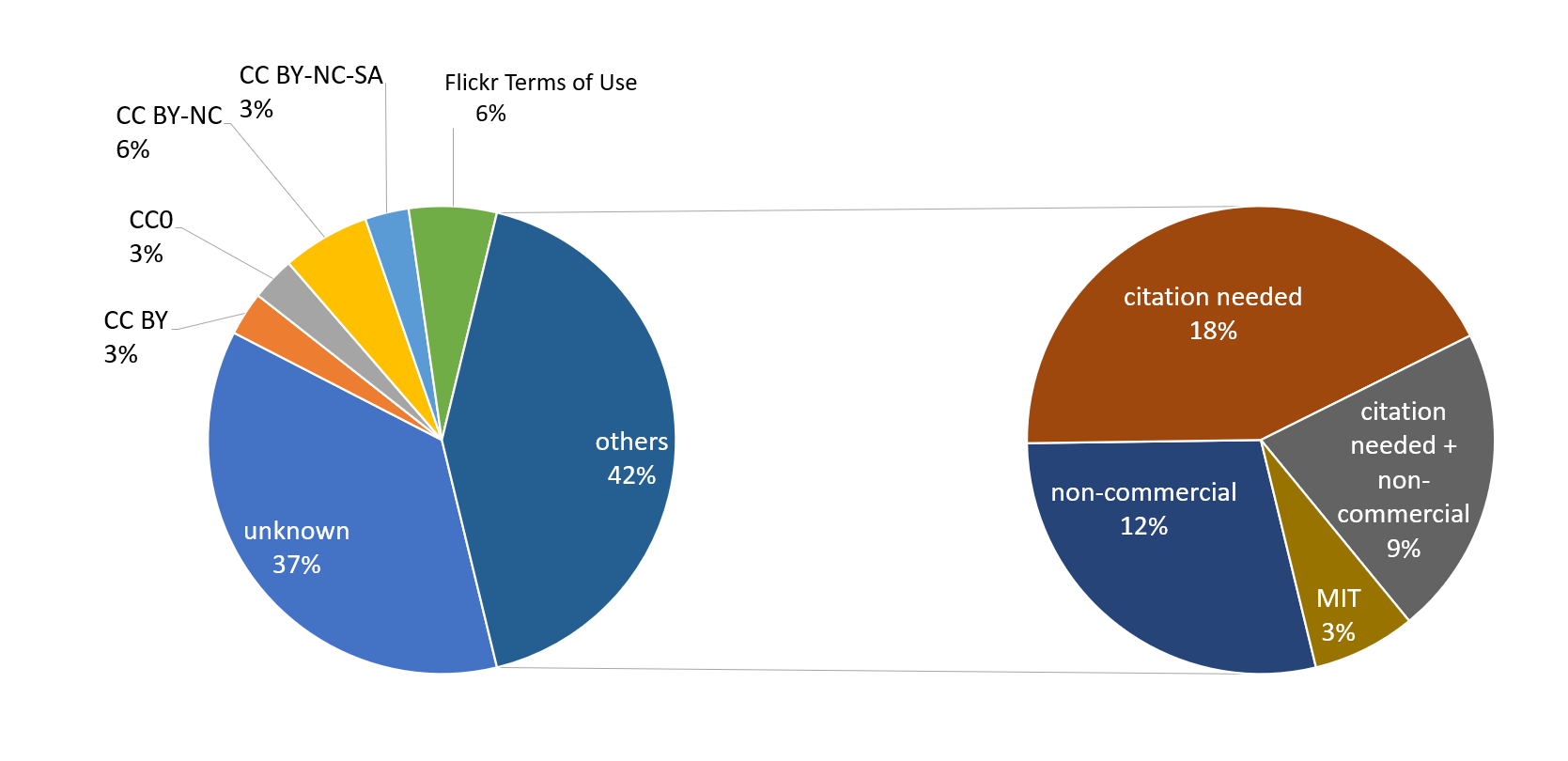}
    \caption{The result of the license of the datasets of Investigation B}
    \label{fig:collectionB}
\end{figure*}

The results we obtained are shown in Figure \ref{fig:collectionB}.

Comparing the pie chart with the results of Investigation A, we can see that the percentage of datasets used in GitHub repositories with unknown licenses has decreased by 6\%. At the same time, the percentage of datasets with the ``citation needed" statement has increased by 8\%.

As shown in Table 3.3, the licensing of 19 of the 34 datasets is unknown, and the datasets that are explicitly unavailable for commercial use have been reduced from 22 in Investigation A reduced to 12.
\begin{table*}[h]
    \caption{Commercial availability of datasets in Investigation B}
    \centering
    \begin{tabular}{c|c}
    \hline
     Number of commercially available datasets & 3 \\
     \hline
    Number of commercially unavailable datasets &  12\\
     \hline
     Unknown & 19\\
     \hline
    \end{tabular}

    \label{tab:collectionB}
\end{table*}

The comparison between Investigation A and Investigation B shows that the percentage of datasets with non-commercially available licenses in open source platforms, such as GitHub, has decreased, probably because some machine learning developers have consciously refrained from using commercially unavailable datasets. However, in general, ImageNet, COCO, and Pascal VOC are all commercially unavailable datasets, while the majority of repositories use these three datasets.

\subsection{Results of Two Investigations}
By investigating the two collections, we understand the complexity of dataset licensing. We found that much of the dataset licensing is unknown. At the same time, some datasets have multiple licenses. This makes it even more challenging to use the dataset based on the license.

Furthermore, it is possible that the license of the dataset can be changed. However, they are generally not changed because, most of the time, the license is unclear. However, when the license is explicit, it has been changed by the original owner of the image.

\section{Main Investigation}
\subsection{The GQM Model}
It stands to reason that in order to improve the analysis quality and implications, systematic and well-developed research studies should adhere to strict models or protocols to guide and maintain them at all times. In order to reach these above conditions for this main investigation, we used the GQM model, which is the Goal, Question, and Metric paradigm proposed by Basili et al. \cite{banimustafa2018gqm}.GQM was initially designed for data collection and analysis in software engineering research with the following basic idea.
\begin{itemize}
    \item{Data collection and analysis must focus on clear and specific objectives, each of which is classified as a set of questions that can be answered quantitatively, each answered by a number of specific metrics.} 
    \item{The data collected based on the indicators are analyzed to generate answers to the questions and thus achieve the defined objectives.}
\end{itemize}

The above two central ideas of GQM can be seamlessly transformed into the central ideas of investigation and research. In building an investigation process based on the GQM model, the first step is to determine the ultimate goal of this investigation, i.e., what is the ultimate goal of this investigation to clarify and achieve. Afterward, a series of questions are asked based on this ultimate goal, which is mainly used to portray the goal. These questions can also be used as research questions in the investigation. finally, the questions are answered by collecting and analyzing metrics. The metrics needed to answer the questions can be both objective and subjective, but they should all be quantifiable.

\begin{figure}
    \centering
    \includegraphics[width=0.5\textwidth]{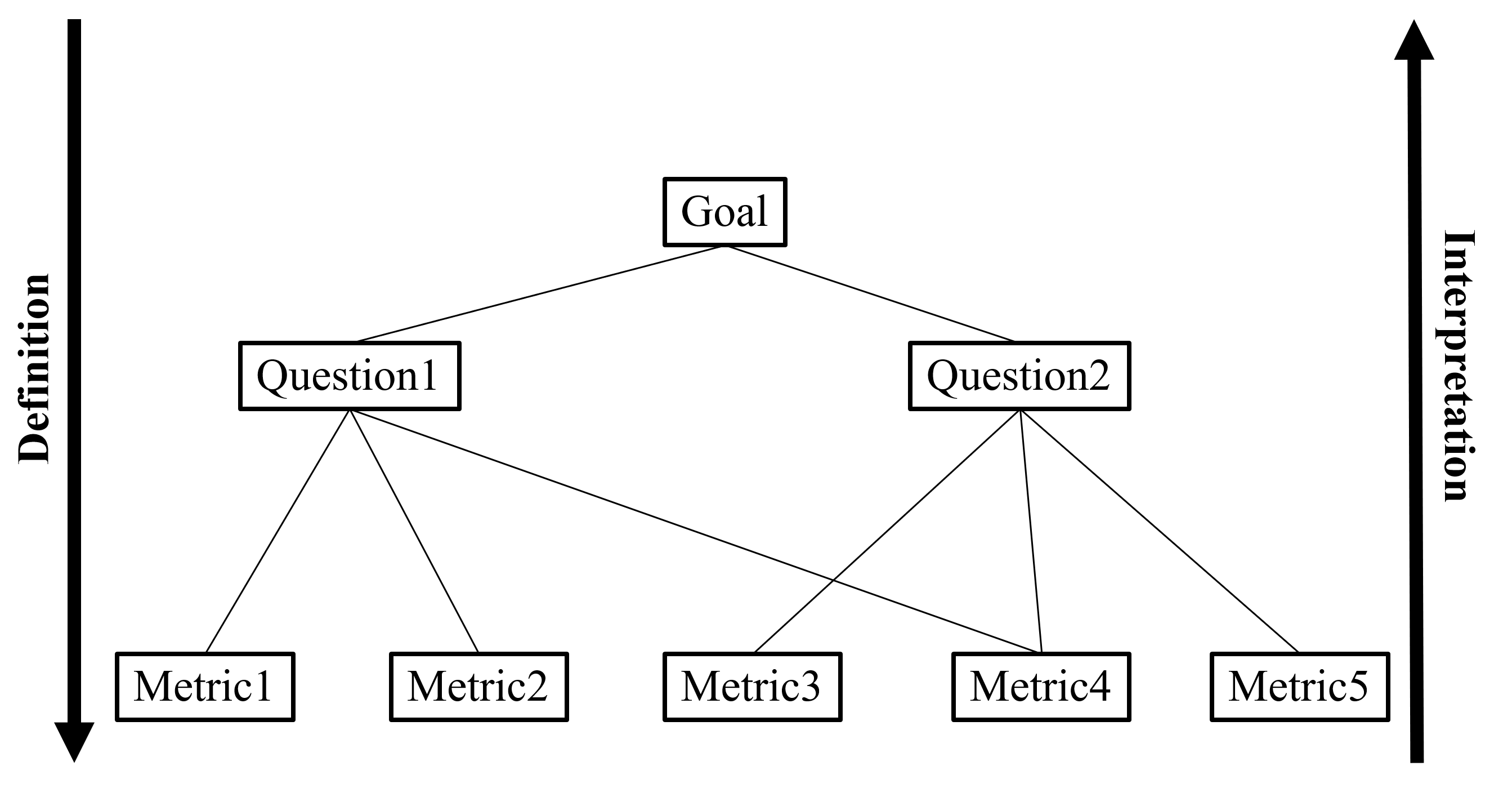}
    \caption{GQM Framework}
    \label{fig:gqmmodel}
\end{figure}

Meanwhile, similar to the field of software engineering, when utilizing GQM, we usually build models by going from top to bottom, i.e., goal to question to metric, and when analyzing and interpreting, we accomplish our ultimate goal by going from bottom to top, i.e., metric to question to the goal(see Figure \ref{fig:gqmmodel}).

\subsection{Goals \& Questions}
In this section, we will introduce each of the two final goals and the research questions. 

In the GQM paradigm of software engineering, GQM models are often constructed through a top-down sequence. However, in this research, since we previously conducted preliminary investigations, we first define research questions, i.e., Q. After that, we then categorize these research questions and analyze how the answers to these questions can help us achieve our goals.
\subsubsection{Goal 1} 
From the two preliminary investigations, we can roughly guess that the licensing system for machine learning datasets is underdeveloped. However, we would like to use this final goal to understand in more detail the licensing content of the datasets, the compatibility of the licensing of the datasets with the data sources, the compatibility of the licensing of the datasets with the systems that use the datasets, and the problems that may occur due to the licensing of the datasets.

\begin{itemize}
    \item [Q1-1]{Are the datasets distributed under commercially available licensing?}\par
    \textbf{Rationale: }We are interested in how much of the extant publicly available dataset is commercially available based on licensing. By interpreting and analyzing this question, we can understand the commercial availability of datasets and analyze the potential for latent licensing violations.
    \item [Q1-2]{Is the licensing compatible between the machine learning systems and the using dataset?}\par
    \textbf{Rationale: }We feel that there is value in studying the use of datasets by machine learning systems nowadays. Machine learning systems often have their own license, and we do not know if this license is compatible with the license of the dataset. By explaining and analyzing this problem, we can know the problems that may have been caused by conflicting licenses.
    \item [Q1-3]{Is the licensing compatible between the dataset and the source data?}\par
    \textbf{Rationale: }Similar to Q1-2, we want to know if the dataset and the license of the data it contains are compatible. Our two preliminary investigations found that many datasets did not indicate their data sources, and the licensing of the data sources may often affect the licensing of the dataset. By explaining this, we can know whether the license of the dataset itself already has the possibility of a license violation.
    \item [Q1-4]{What problems could occur if the dataset is not in accordance with its licensing?}\par
    \textbf{Rationale: }In the last problem of G1, we put our eyes on the problems that can occur by not complying with the license of the dataset. We want to know what problems have occurred or what problems are possible up to now. By analyzing and interpreting this question, we can understand the importance of dataset licensing and the severity of violating dataset licensing, helping us to make recommendations to the dataset community accordingly.
\end{itemize}

\subsubsection{Goal 2}
In Goal 2, our primary goal is to find out whether the existing dataset is easy to use based on the license. Specifically, what we can do with the datasets based on licensing, what changes we can make to the datasets, what impact these changes have on the datasets, and also whether there are tools to help machine learning system developers to use the datasets.

\begin{itemize}
    \item [Q2-1]{Under what licensing are the (image) datasets for machine learning distributed?}\par
    \textbf{Rationale: }To understand how easy it is to use a license-based dataset, we must first understand how the existing publicly available datasets are usually distributed with a license. By analyzing and explaining this issue, we can facilitate the organization of the usability of these datasets.
    \item [Q2-2]{What problems have been caused as a result of changes to datasets for machine learning?}\par
    \textbf{Rationale: }With the aim of understanding the ease of use of the dataset, we were interested in the problems that arise due to the changes that occur in the dataset. We wanted to know to what extent changes to the dataset would affect the machine learning system using it. Analyzing and interpreting this problem enables us to understand the problems that may arise due to changes in the dataset.
    \item [Q2-3]{Is there any management based on licensing of datasets?}\par
    \textbf{Rationale: }We are interested in existing dataset management tools. We would like to know what kind of features are generally available in existing dataset management tools and, most importantly, whether management tools based on dataset licensing are available. By analyzing and answering this question and summarizing the two goals, we can get a new direction for development and research.
\end{itemize}

Therefore, the second goal we set based on the above three research questions is: \textbf{\emph{Is it easy to use machine learning datasets following the licensing?}}

\subsection{Metrics}
\begin{itemize}
    \item [M1.]{\textbf{Type of license for the dataset.}}
    We organized the license types of the dataset and defined them categorically. This metric is the cornerstone of all our metrics, and the information and definitions of licenses will be derived from the classification of licenses. This metric will help to explain Q1.1, Q1.2, Q1.3, and Q2.1.
    \item [M2.]{\textbf{Commercial availability of each license.}}
    We will conduct a commercial availability survey for each license type. The commercial availability of each license allows us to know about potential license violations. This metric will help to explain Q1-1 and Q2-1.
    \item [M3.]{\textbf{Source of the original data in the dataset.}}
    We will investigate the data sources of the dataset and analyze the licensing of these data sources. The complexity of the dataset is verified by the original dataset, which facilitates our investigation of the licensing of the original data. This metric will help to explain Q1-3.
    \item [M4.]{\textbf{Number of datasets with explicit licenses.}}
    We will count the number of datasets with explicit licenses. The number of explicit datasets gives us an idea of how well-established and valued the licensing system is for the datasets. This metric will help to explain Q2-1.
    \item [M5.]{\textbf{Number of datasets with multiple licenses.}}
    Datasets with multiple licenses are complicated to use and need to be clarified. By being explicit about this metric, we can grasp the complexity of using the dataset. This metric will help to explain Q1-1 and Q2-1. 
    \item [M6.]{\textbf{Number of datasets with licensing issues.}}
    We will investigate what kind of problems occur in the dataset due to license issues and the number of datasets where problems occur. By investigating the licensing issues occurring in the dataset, we can clarify how the problems occur in the dataset and facilitate our summary and generalization. This metric will help to explain Q1-4 and Q2-2.
    \item [M7.]{\textbf{Number of systems using the withdrawn dataset.}}
    We will search GitHub and Google for existing systems still using the withdrawn dataset. The statistics of this metric facilitate us in organizing what kind of impact on the system will be caused when the dataset is affected. This metric will help to explain Q2-2. 
    \item [M8.]{\textbf{Usage of datasets in repositories.}}
    We will investigate GitHub for machine learning repositories and the datasets they use. This metric is also one of our cornerstones. We compiled our research by analyzing the most frequently used datasets on GitHub. This metric will help to explain Q1-2 and Q2-3. 
    \item [M9.]{\textbf{Status of the system's own license.}}
    We will check the license of M8 repositories to see if it is compatible with the license of the dataset. We need to know the license of the machine learning system in order to determine whether the system is compatible with the license of the dataset. This metric will help to explain Q1-2.
    \item [M10.]{\textbf{Number of incompatible licenses for the image itself and the dataset.}}
    We will check the license of the image from the source and investigate whether it is compatible with the license of the dataset. After understanding the data source of the dataset, we need to clarify whether the license to get the dataset is compatible with the data it contains. This metric will help to explain Q1-3. 
    \item [M11.]{\textbf{Number of systems affected by dataset licensing issues.}}
    We will investigate what kind of problems can occur in machine learning systems due to changes in the dataset and what the impact can be. We hope to investigate this metric to learn precisely how the issue of dataset licensing affects the system. This metric will help to explain Q2-2.
    \item [M12.]{\textbf{Number of supporting tools on the dataset.}}
    We will search GitHub for tools related to dataset management and inquire about their functionality. We need to understand the number of dataset management tools in existence, as well as their functionality. This metric will help to explain Q2-3.
\end{itemize}

Figure \ref{GQM1} in Appendix shows the mapping relations between Goal 1 and the contained question and metrics, while Figure \ref{GQM2} in Appendix shows the mapping relations under Goal 2. 

\subsection{Results}
In this section, we will show the findings and results of the 12 metrics. We have to make the point that all of these investigations were conducted between April and December 2022.

\noindent\textbf{M1: Type of license for the dataset.}\par
Through the two preliminary investigations, as well as the investigation of datasets used by machine learning systems on GitHub mentioned in the following metrics, we have organized and categorized the types of dataset licenses. Among the datasets with ``known" licenses, we classified the dataset licenses into 8 categories. They are citation needed, non-commercial, Flickr Terms of Use, CC0, CC BY, CC BY-NC, MIT, and CC BY-NC-SA. The definitions of citation needed, non-commercial, and Flickr Terms of Use are explained in our Investigation A(read Section 4.1).

Among these categories, citation needed and non-commercial are the categories of licenses for datasets that declare either of these requirements. If we find a dataset that has both an explicit license, such as CC BY-NC, and a non-commercial statement, we will only classify the license of the dataset under the explicit license, i.e., CC BY-NC.

\noindent\textbf{M2: Commercial availability of each license.}\par
We investigated the commercial availability of licenses for the dataset summarized by M1 and obtained the results shown in Table \ref{tab:commerciallylicenses}.
\begin{table*}[h]
    \caption{Number of commercially available licenses}
    \centering
    \begin{tabular}{c|c}
    \hline
     Commercially available    & 3 \\
     \hline
     Commercially unavailable    & 3 \\
     \hline
     Unknown    & 2 \\
     \hline
    \end{tabular}

    \label{tab:commerciallylicenses}
\end{table*}

In this classification, the commercial availability of datasets with explicit licenses is not required for us to define, for example, commercial availability for MIT as possible. In contrast, in our own definition of citation needed, non-commercial, and Flickr Terms of Use, we define citation needed and Flickr Terms of Use as unknown and non-commercial as commercially unavailable.

As discussed here, it is generally accepted that the citation needed is within the scope of commercial availability, but since the dataset does not have a complete licensing system, we define the commercial availability of the citation needed as unknown. For Flickr Terms of Use, these datasets all use Flickr as the data source. We asked the owners of these datasets via email in May 2022 if they used a specific license of the Flickr API\footnote{https://www.flickr.com/services/api/
} when acquiring the images, but only one email was returned telling us that the data was collected without a specific license for the image source.

\noindent\textbf{M3: Source of the original data in the dataset.}\par
Table \ref{tab:m3} shows the data sources of the 42 datasets we investigated. 
\begin{table*}[h]
    \caption{Data source status of the dataset}
    \centering
    \begin{tabular}{c|c|c}
    \hline
    Source & Initial results & Results after asking \\
    \hline
     Unknown    & 22 &  21 \\
    \hline
     Developers    & 9 &  10\\
    \hline
      Flickr   & 4 & 4 \\
    \hline
      NPO   &3  &3 \\
    \hline    
      Other datasets  & 4 &4\\
    \hline    
    \end{tabular}

    \label{tab:m3}
\end{table*}

We need to interpret our two columns of results. The results we obtained in our investigation up to November 2, 2022, are shown in the Initial results of Table \ref{tab:m3}. However, since it was found that the origin of most of the datasets in the survey was not clear, we decided to send an email to the dataset developers who could find contact information on the dataset homepage, asking about the origin of the datasets and using this opportunity to ask about the commercial possibilities of datasets whose licenses were otherwise unknown. As a result, we sent 14 emails, and as of December 14, 2022, we have received 12 replies, and the changes in our findings as a result of these replies will be indicated in the results that follow. In M3, our results are shown in Results after asking, based on the results of the returned emails.

Based on our investigation, we divided the data sources into five categories: unknown, developer, Flickr, NPO, and others. Unknown means that the source of the dataset is unspecified or that the images were obtained from sources such as the Internet or Google; Developer means that the dataset was made from images owned by the dataset developer; Flickr means that the images in the dataset were obtained from Flickr; NPO means that the data in the dataset were obtained in an NPO project. Other datasets indicate that the dataset is a subset of another dataset or a modified product of another dataset.

\noindent\textbf{M4: Number of datasets with explicit licenses.}\par
Table \ref{tab:m4} shows the licensing status of the 42 datasets. 
\begin{table*}[h]
    \caption{Number of datasets with explicit licenses}
    \centering
    \begin{tabular}{c|c|c}
    \hline
    Type & Initial results & Results after asking \\
    \hline
     Unknown    & 14 &  11 \\
    \hline
     Flickr Terms of Use    & 4 &  4\\
    \hline
      CC Series or MIT   & 9 & 9 \\
    \hline
      Citation needed   & 6  &5 \\
    \hline    
      Non-commercial  & 4 &6\\
    \hline  
     Citation needed + Non-commercial    & 5 &  7 \\  
    \hline
    \end{tabular}

    \label{tab:m4}
\end{table*}

In our classification, citation needed, non-commercial, and citation needed + non-commercial are divided into three categories. We can learn from the table that four datasets become commercially unavailable from the case of unknown commercial availability based on the return letters.

\noindent\textbf{M5: Number of datasets with multiple licenses.}\par
In our investigation, we found two datasets with multiple licenses in opposite situations: ADE20K\footnote{https://groupsimpliesl.mit.edu/vision/datasets/ADE20K[Last visited:2022/12/01]} and FFHQ\footnote{https://github.com/NVlabs/ffhq-dataset[Last visited:2022/12/01]}.In ADE20K, it declares that the image is non-commercial; however, it declares that for the annotation of the image or something like that, it has a BSD-3-Clause license, i.e., commercially available. Whereas in FFHQ, we can find that the image declaration is Flickr Terms of Use, and for annotations, the license they declare is CC BY-NC-SA 4.0, i.e., commercially unavailable.

Due to the imperfection of our own dataset licensing system and the scarcity of datasets with explicit licenses, we could only find these two datasets with opposite scenarios.

\noindent\textbf{M6: Number of datasets with licensing issues.}\par
We investigate this metric through the aggregation of Adam et al. \cite{Exposing.ai} for the dataset with licensing issues. We obtained seven datasets where problems occurred due to licensing, they are MS-Celeb-1M, Tinyimage, DukeMTMC, VGGface, UCCS, and Brainwash.

Through our investigation, we learned that all seven datasets were withdrawn, which means they are not publicly available on the Internet anymore. At the same time, they were withdrawn for reasons of privacy, harm caused by annotation, and ethically questionable purposes. The licenses for these datasets also do not declare anything relevant. Due to the incompleteness of the licensing system, datasets are often withdrawn from public availability when they are criticized.

\noindent\textbf{M7: Number of systems using the withdrawn dataset.}\par
Table \ref{tab:m7} shows the findings for systems that are still using the withdrawn dataset.
\begin{table*}[h]
    \caption{Number of systems using the withdrawn dataset}
    \centering
\begin{tabular}{l|l|l}
\hline
Dataset     & \begin{tabular}[c]{@{}l@{}}Number of machine learning \\ systems\end{tabular} & \begin{tabular}[c]{@{}l@{}}Number of all \\ searchable repositors\end{tabular} \\ \hline
MS-Celeb-1M & 4                                                                             & 13                                                                             \\ \hline
VGGface     & 5                                                                             & 7                                                                              \\ \hline
Tiny image  & 2                                                                      ; they & 2                                                                              \\ \hline
DuckMTMC    & 2                                                                             & 12                                                                             \\ \hline
Megaface    & 8                                                                             & 15                                                                             \\ \hline
UCCS        & 0                                                                    & 0                                                                              \\ \hline
Brainwash   & 4                                                                    & 5                                                                              \\ \hline
\end{tabular}

    \label{tab:m7}
\end{table*}

We first searched through GitHub with the names of the seven datasets as keywords, and we could only get the dataset cleaning-related repository. We could not find any machine learning system that had used the dataset. So, we searched through Google with ``dataset name" + ``GitHub" and got the information shown in Table \ref{tab:m7}. The third column of the table shows the number of repositories we can search for this dataset. In contrast, the second column indicates the number of repositories of machine learning systems that use this dataset.

One point worth raising is that in the study of Peng et al. \cite{mitigating}, they found 21 projects on GitHub still using this dataset after MS-Celeb-1M was withdrawn. Still, on December 10, 2022, only four projects were using the dataset, according to our investigation.

\noindent\textbf{M8: Usage of datasets in repositories.}\par
With the two preliminary investigations, we decided to expand the main survey with GitHub-related topics. The investigation results are shown in Table \ref{tab:m8repository}.

We searched through GitHub's advanced search for repositories in topics related to visual machine learning systems. we chose the top 70 repositories with the highest number of stars in each related topic as the scope of the investigation, i.e., a total of 350 repositories. After eliminating the repositories of the collection and summary, 311 repositories were left. And after eliminating the repositories that may repeatedly appear in each topic, we have 263 repositories left. Among these 263 systems, 210 states have a license, 116 that explicitly state what dataset they use, and 70 declare both their license and the dataset they use.

\begin{table*}[h]
 \caption{Licensing and dataset usage of the repository}
 \centering
\begin{tabular}{l|l|l|l}
\hline
Topic                                                           & \begin{tabular}[c]{@{}l@{}}Number of \\ repositories \\ holding \\ explicit licenses\end{tabular} & \begin{tabular}[c]{@{}l@{}}Number of \\ repositories \\ that show \\ the datasets used\end{tabular} & \begin{tabular}[c]{@{}l@{}}Number of \\ repositories that\\ have both the\\ first two features\end{tabular} \\ \hline
\begin{tabular}[c]{@{}l@{}}image\\ classification\end{tabular}  & 49                                                                                                & 38                                                                                                  & 27                                                                                                          \\ \hline
\begin{tabular}[c]{@{}l@{}}image\\ annotation\end{tabular}      & 43                                                                                                & 20                                                                                                  & 17                                                                                                          \\ \hline
\begin{tabular}[c]{@{}l@{}}object\\ detection\end{tabular}      & 53                                                                                                & 39                                                                                                  & 29                                                                                                          \\ \hline
recognition                                                     & 54                                                                                                & 16                                                                                                  & 10                                                                                                          \\ \hline
\begin{tabular}[c]{@{}l@{}}semantic\\ segmentation\end{tabular} & 46                                                                                                & 25                                                                                                  & 14                                                                                                          \\ \hline
SUM                                                             & 245                                                                                               & 138                                                                                                 & 97                                                                                                          \\ \hline
\end{tabular}
\label{tab:m8repository}
\end{table*}
In Table \ref{tab:m8repository}, we represent the original investigation results data under each topic, i.e., the number of repositories that have not been eliminated as duplicates in each topic. The first column in Table \ref{tab:m8repository} shows the name of each topic, the second column shows the number of repositories in each topic that have declared their own licenses, and the third column shows the number of repositories in each topic that have declared what datasets they use, and the fourth column shows the number of repositories that have declared both their own licenses and what datasets they use.

Table \ref{tab:m8dataset} shows the datasets used in all the investigated machine learning systems.
\begin{table*}[h]
\caption{Usage of each dataset}
\centering
\begin{tabular}{l|l}
\hline
Dataset                                                                                            & Number of the repositories \\ \hline
ImageNet                                                                                           & 35                         \\ \hline
COCO                                                                                               & 21                         \\ \hline
Pascal VOC                                                                                         & 20                         \\ \hline
MINIST                                                                                             & 13                         \\ \hline
CIFAR                                                                                              & 10                         \\ \hline
FFHQ                                                                                               & 9                          \\ \hline
ADE20K                                                                                             & 8                          \\ \hline
Caltech                                                                                            & 6                          \\ \hline
Cityspaces                                                                                         & 5                          \\ \hline
\begin{tabular}[c]{@{}l@{}}...(The repository \\ that uses the dataset \\ is under 5)\end{tabular} & ...                        \\ \hline
SUM                                                                                                & 184                        \\ \hline
\end{tabular}
\label{tab:m8dataset}
\end{table*}

We know from the table that the number of datasets the repositories explicitly use is 116, while their sum is 184. In other words, we found an average of 1.5 datasets used per system. It is also worth pointing out that the dataset we compiled here is the same one that was investigated in M4 above.

\noindent\textbf{M9: Status of the system's own license.}\par
In this metric, we investigated the commercial availability of the repository's license and analyzed 70 repositories that both indicated their own license and the dataset used. According to the investigation of the license of the repository, we got 9 kinds of licenses: MIT, Apache-2.0, BSD-3-Clause, GPL-3.0, AGPL-3.0, LGPL-3.0, CC0, WTFPL license, and CC BY-SA. These licenses are all commercially available.

\begin{table*}[h]
    \caption{License compatibility of datasets and systems}
    \centering
\begin{tabular}{l|l|l}
\hline
             & Initial results & Results after asking \\ \hline
Compatible   & 10              & 10                   \\ \hline
Incompatible & 29              & 31                   \\ \hline
Unknown      & 31              & 29                   \\ \hline
\end{tabular}

    \label{tab:m9}
\end{table*}
Table \ref{tab:m9} shows the licensing compatibility of the 70 machine learning systems we investigated and the datasets they used. While the licenses of the 70 repositories are all commercially available, we consider license incompatibility cases where the licenses of the datasets are commercially unavailable. In contrast, a combination of system and dataset is considered compatible if the dataset’s license is commercially available. Moreover, we treat compatibility as unknown if the license of the dataset used by the system is unknown.

From Table \ref{tab:m9}, almost half of the systems actually have licensing conflicts with the licenses of the datasets it uses. Furthermore, the commercial availability of licenses for datasets is primarily unknown, resulting in the licensing compatibility of systems and datasets being mainly unknown.

\noindent\textbf{M10: Number of incompatible licenses for the image itself and the dataset.}\par
The first row of Table \ref{tab:m10} refers to the number of datasets that explicitly indicate the source of the data in the dataset, and the second row refers to the number of datasets that are compatible with the license of the data it contains.

\begin{table*}[h]
    \caption{License conflict situations for datasets and its data}
    \centering
\begin{tabular}{l|l|l}
\hline
                                                                                        & Initial results & Results after asking \\ \hline
\begin{tabular}[c]{@{}l@{}}Number of datasets \\ with explicit data source\end{tabular} & 20              & 21                   \\ \hline
Compatible                                                                              & 11              & 12                   \\ \hline
\end{tabular}

    \label{tab:m10}
\end{table*}

In the results of our investigation, we found that datasets and the data licenses they contain are primarily compatible and commercially available only for datasets where the data originates from NPOs. Other than that, datasets, where the data originates from itself, are mostly commercially unavailable.

\noindent\textbf{M11: Number of systems affected by dataset licensing issues.}\par
For this metric, we know that in the preliminary investigations of the repository in GitHub, 6 out of 116 machine learning systems are still using datasets that have been withdrawn. However, we consider this to be insufficient.

Due to the absence and incompleteness of the dataset licensing management system and the penalty system for non-compliance, it can only be summarized by others' opinions on the possible impact of datasets on the system.

Even after the dataset is withdrawn, the dataset can still be found on the Internet, and the dataset can still potentially be used by machine learning systems \cite{mitigating}. A great deal of machine learning systems using the withdrawn dataset has been criticized by the academic community \cite{dirtysecret, Largeimagedatasets, contractor2022behavioral}. However, in the application community of machine learning systems, or instead the industry, perceptions related to the disappearance of datasets due to licensing of datasets to the unavailability of the system have proliferated \cite{causingadecay, currentAi}.

With an as-yet-improved dataset licensing system, we have no way to determine precisely how the dataset will affect the system, but we can be sure that the impact will be significant.

\noindent\textbf{M12: Number of supporting tools on the dataset.}\par
We searched on GitHub using the dataset tool as a keyword. We selected the first 40 repositories in order of the most stars. 21 of these repositories are tools related to machine learning datasets. These tools have features such as data augmentation, help labeling datasets, adding and deleting datasets, etc. However, among the 21 dataset tools, none of them are license-based to help manage datasets for machine learning systems.

\section{Discussion}
In this section, we will first introduce our answer to the question of the GQM model. Afterward, We will discuss some recommendations for the machine learning community and the developers of machine learning datasets.

\subsection{Answer to Research Questions}
\subsubsection{Goal 1}
In this section, we will solve the research question under Goal1: \textbf{Are there licensing issues in the machine learning dataset?} \par
\noindent\textbf{[Q1-1]Are the datasets distributed under commercially available licensing?} \par
\noindent\textbf{Answer: }Among the classified licensing categories, Non-commercial, CC BY-NC, and CC BY-NC-SA are explicitly not available for commercial use, while CC0, CC BY, and MIT are explicitly allowed for commercial use. However, we found that the datasets used by most machine learning systems are generally commercially unavailable. 

\noindent\textbf{[Q1-2]Is the licensing compatible between the machine learning systems and the using dataset?} \par
\noindent\textbf{Answer: }Only 10 cases in which the license of the machine learning system and the dataset it uses are compatible, while the cases that are incompatible are 29. Therefore, it is inferred that the commercial availability of most machine learning systems is incompatible with the using dataset.

\noindent\textbf{[Q1-3]Is the licensing compatible between the dataset and the source data?} \par
\noindent\textbf{Answer: }It is possible that the license of the dataset and the images contained in this dataset is different. Also, each image contained in the dataset may have a different license. Furthermore, we found that about half of the datasets are incompatible with the images it contains for commercial availability.

\noindent\textbf{[Q1-4]What problems could occur if the dataset is not in accordance with its licensing?} \par
\noindent\textbf{Answer: }Violated datasets could be withdrawn. Currently, the emphasis is on the license of the dataset used in machine learning. The annotation of the dataset may offend others. Failure to obtain the appropriate license may violate the law and lead to the withdrawal of the dataset.

After explaining the above four questions, we can come to accomplish our Goal 1: \textbf{Are there licensing issues in the machine learning dataset?} 

\begin{center}
\noindent\fbox{
\parbox{.9\linewidth}{
The result of Goal 1: Based on our research on the commercial availability of dataset licensing, we analyze the potential licensing conflicts between the dataset and the data it contains and the potential licensing violations that could arise between the dataset and the machine learning systems that use it. Based on the results of the above investigation, it is found that the licensing of a machine learning dataset is affected by the licensing of the data it contains. The licensing of the dataset affects the commercial availability of the system using the dataset. When problems occur with the licensing of a dataset, it can lead to the withdrawal of the dataset and thus affect the development of machine learning systems.}
}
\end{center}
\subsubsection{Goal 2}
In this section, we will solve the research question under Goal2: \textbf{Is it easy to use machine learning datasets following the licensing?} \par
\noindent\textbf{[Q2-1]Under what licensing are the (image) datasets for machine learning distributed?} \par
\noindent\textbf{Answer: }Through the investigation of 42 datasets, we summarized seven types of licenses, namely: Citation needed; Non-commercial; Flickr Terms of Use; the four CC series licenses, CC0, CC BY, CC BY-NC, CC BY-NC-SA; and MIT. In addition, 11 out of the 42 datasets are not explicitly labeled with their own license.

\noindent\textbf{[Q2-2]What problems have been caused as a result of changes to datasets for machine learning?} \par
\noindent\textbf{Answer: }We found that some machine learning systems still use withdrawn datasets, but the number of such machine learning systems is decreasing. However, due to the lack of management of the dataset licensing, changes to datasets do not significantly impact the use of datasets by machine learning systems.

\noindent\textbf{[Q2-3]Is there any management based on licensing of datasets?} \par
\noindent\textbf{Answer: }We searched for dataset-related tools through Google and GitHub and found many tools to support the management of datasets, including data augmentation, annotation, and modification. However, we found no tool to support licensing management of datasets.

After explaining the above four questions, we can come to accomplish our Goal 2: \textbf{Is it easy to use machine learning datasets following the licensing?}

\begin{center}
\noindent\fbox{
\parbox{.9\linewidth}{
The result of Goal 2: Based on our summary of the types of existing dataset licenses, our compilation of the issues raised by dataset licensing problems, and our survey of publicly available dataset management tools, we can explain the state of dataset usage.  Based on the current conditions where dataset licensing is not complete, it is easy to use datasets due to the lack of tracking management for datasets, and even if a dataset is withdrawn, it can still be found through the leftover files. However, due to the lack of management of license-based datasets, there is no method used to reduce the risk of licensing violations in machine learning systems.}
}
\end{center}

\subsection{Recommendations}
Based on the investigations in this thesis, we present here our recommendations to AI engineers who use datasets, data scientists who develop and create datasets, and the whole AI community, respectively.
\begin{itemize}
    \item{For AI engineers}\par
    \textbf{Recommendations1: Choose the dataset that has the proper license according to the purpose of the machine learning system you are developing \cite{cani}.} While many datasets are now publicly available with unclear licensing, when developing machine learning systems for commercial use, be sure to confirm that the dataset you are using is commercially available.\par
    \textbf{Recommendations2: If you are using a project on an open-source site for commercial purposes, you should be aware of the type of license for the dataset of this project \cite{dirtysecret}.}Based on the investigation, we can know that a portion of the repositories uses datasets that are incompatible with the system's license. To avoid licensing violations, engineers should be aware of both the system and the dataset being utilized.\par
    \item{For data scientists}\par
    \textbf{Recommendations3: When collecting data, try to ensure that the data source is as homogeneous as possible \cite{2019towardstraceability,2021towardsaccountability}}. With a single data source, you can confirm that you are in control of the source of your data. And after specifying the data source, you can clearly know the permission of the metadata of the data source. When a problem occurs with the data in the data source, the developer of the dataset can immediately take measures to modify the dataset.\par
    \textbf{Recommendations4: Select an appropriate license when releasing the dataset \cite{mitigating}.}In order to facilitate the use of datasets by machine learning systems and to manage dataset licensing, it is necessary for dataset developers to choose appropriate and formal (not ``citation needed" statements) licenses based on the source of the data when releasing.\par
    \item{For community}\par
    \textbf{Recommendations5: Improve accountability mechanisms for \\datasets \cite{towardsmon, 2021towardsaccountability}.}Accountability mechanisms for the datasets of machine learning systems are essential. However, accountability mechanisms are needed for the entire machine learning community. Accountability mechanisms for datasets can modernize the AI development process and allow it to evolve toward an ethical framework.\par 
\end{itemize}

\section{Conclusion}
In this section, we will conclude the whole thesis and discuss future work.

\subsection{Conclusion}
The role of datasets in AI system development is critical due to the large-scale development and reliance on machine learning systems. However, we identified possible licensing violations in the use of the dataset by the machine learning system. To clarify and address this situation, we conducted a systematic investigation of the licensing of datasets.

First, we clarify the possible problems of improper dataset licensing and unknown commercial availability in machine learning systems through the two preliminary investigations. The emergence of these problems rings alarm bells about the use of non-commercial datasets in commercial machine learning systems.

Subsequently, we conducted a systematic investigation of the dataset, the data in the dataset, and the system using the dataset based on the GQM model. After getting the results of the investigation, we discussed the results and got the answers to 7 research questions. Afterward, we provided recommendations to AI engineers, dataset developers, and all dataset-related communities.

\subsection{Future Work}
After a systematic investigation of the licensing status of the dataset, two directions for future research can be extended.
\begin{itemize}
    \item{Automatically evaluating dataset licensing is difficult because the accountability system for datasets is not yet complete. Once the accountability mechanisms are complete, studies to automatically assess the correctness of dataset licensing can be discussed.}
    \item{After conducting the Investigation A and Investigation B, we developed a DVC\footnote{https://dvc.org}-based dataset query system. Since no automated evaluation system for dataset licensing exists, the datasets in this system are entered manually by us. When a mature and complete automatic evaluation system for datasets is available, we can develop a management tool based on dataset licensing.}
\end{itemize}

\clearpage
\onecolumn

\section*{Appendix}
\begin{figure}[h]
    \begin{center}

    \centering
    \includegraphics[width=0.78\columnwidth]{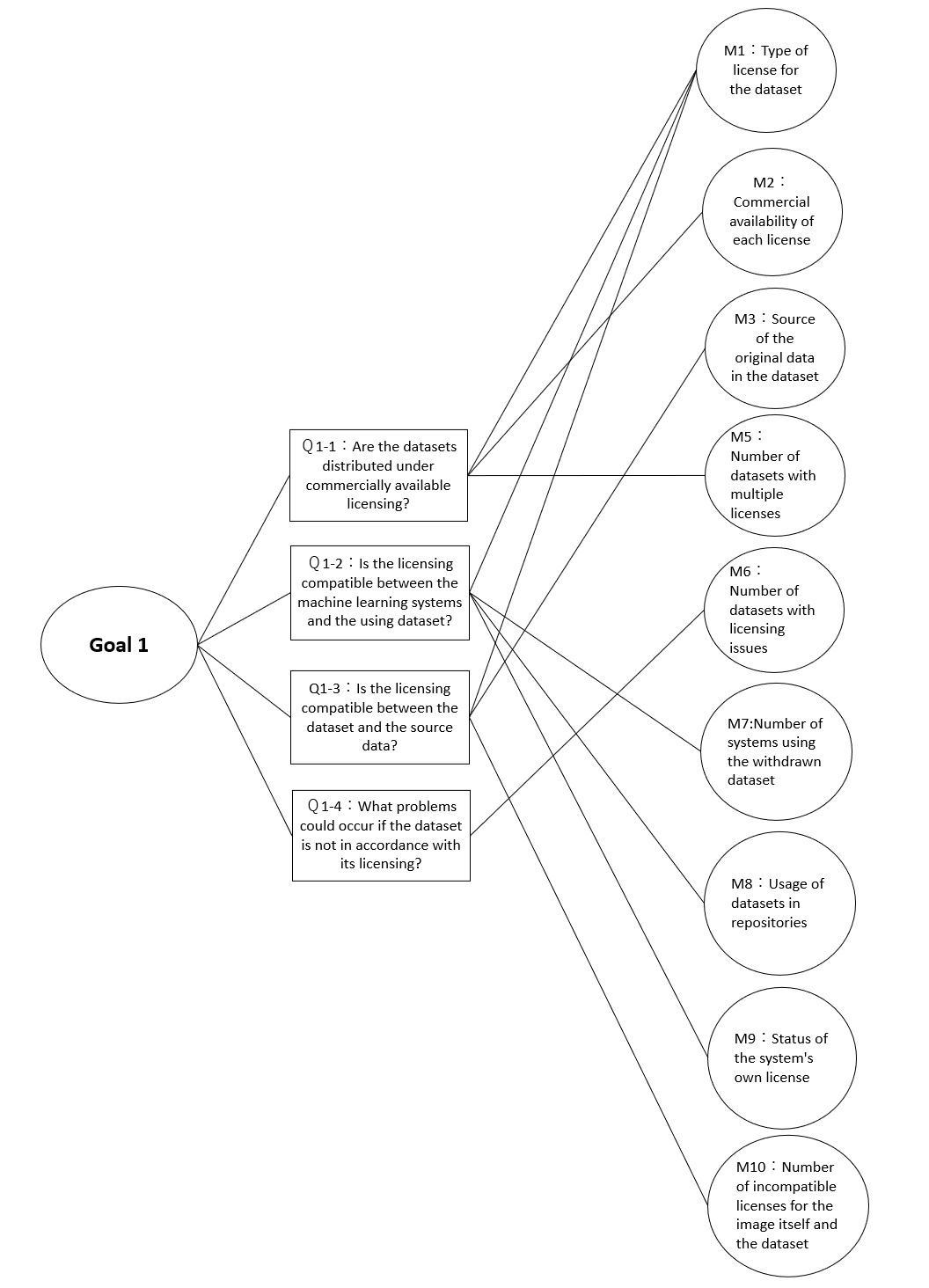}
    \caption{The mapping relationship between Q and M in G1}
    \label{GQM1}
    \end{center}
    \end{figure}

\begin{figure}[h]
    \centering
   \includegraphics[width=0.78\columnwidth]{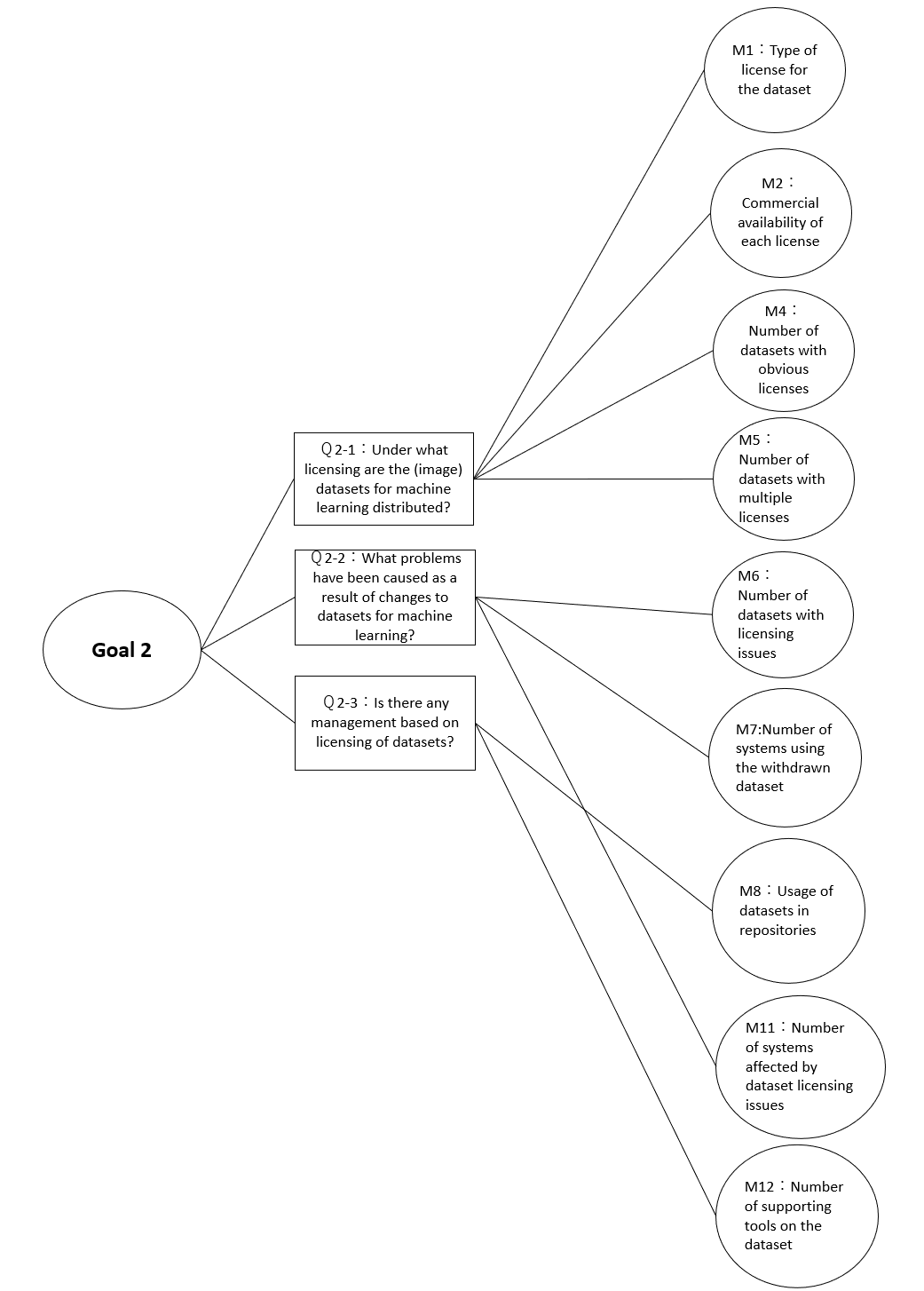}
    \caption{The mapping relationship between Q and M in G2}
    \label{GQM2}
\end{figure}

%\twocolumn[
%    \appendix
As mentioned in Section IV.C, Figure \ref{GQM1} and Figure \ref{GQM2} illustrate the mapping relationships between Questions and Metrics in Goal1 and Goal2.

%    \caption{The mapping relationship between Q and M in G1}
%    \end{center}
 
%    \section{Relationship mapping diagram of GQM}
%    \subsection{The mapping relationship between Q and M in G1}
%    \label{GQM1}
%]
%    ]
%\clearpage

%\twocolumn[
%\subsection{The mapping relationship between Q and M in G2}
%\label{GQM2}
%]
%]

\vspace{12pt}

\begin{thebibliography}{10}

\bibitem{amershi2019software}
Saleema Amershi, Andrew Begel, Christian Bird, Robert DeLine, Harald Gall, Ece
  Kamar, Nachiappan Nagappan, Besmira Nushi, and Thomas Zimmermann.
\newblock Software engineering for machine learning: A case study.
\newblock In {\em 2019 IEEE/ACM 41st International Conference on Software
  Engineering: Software Engineering in Practice (ICSE-SEIP)}, pages 291--300.
  IEEE, 2019.

\bibitem{currentAi}
Martin Anderson.
\newblock {\em Current AI Practices Could Be Enabling a New Generation of
  Copyright Trolls}.
\newblock [Last visited: 2022/12/09].

\bibitem{banimustafa2018gqm}
Ahmed BaniMustafa.
\newblock Predicting software effort estimation using machine learning
  techniques.
\newblock In {\em 2018 8th International Conference on Computer Science and
  Information Technology (CSIT)}, pages 249--256. IEEE, 2018.

\bibitem{2019towardstraceability}
Iain Barclay, Alun Preece, Ian Taylor, and Dinesh Verma.
\newblock Towards traceability in data ecosystems using a bill of materials
  model.
\newblock In {\em International Workshop on Science Gateways}. CEUR-WS, 2019.

\bibitem{towardsmon}
Misha Benjamin, Paul Gagnon, Negar Rostamzadeh, Chris Pal, Yoshua Bengio, and
  Alex Shee.
\newblock Towards standardization of data licenses: The montreal data license.
\newblock {\em arXiv preprint arXiv:1903.12262}, 2019.

\bibitem{Largeimagedatasets}
Abeba Birhane and Vinay~Uday Prabhu.
\newblock Large image datasets: A pyrrhic win for computer vision?
\newblock {\em 2021 IEEE Winter Conference on Applications of Computer Vision
  (WACV)}, pages 1536--1546, 2021.

\bibitem{breiman2001random}
Leo Breiman.
\newblock Random forests.
\newblock {\em Machine learning}, 45(1):5--32, 2001.

\bibitem{breueldiscipline}
Cristiano Breuel.
\newblock {\em ML Ops: Machine Learning as an Engineering Discipline}.
\newblock [Last visited: 2022/12/15].

\bibitem{contractor2022behavioral}
Danish Contractor, Daniel McDuff, Julia~Katherine Haines, Jenny Lee,
  Christopher Hines, Brent Hecht, Nicholas Vincent, and Hanlin Li.
\newblock Behavioral use licensing for responsible ai.
\newblock In {\em 2022 ACM Conference on Fairness, Accountability, and
  Transparency}, pages 778--788, 2022.

\bibitem{coughlan2020standardizing}
Shane Coughlan.
\newblock Standardizing open source license compliance with openchain.
\newblock {\em Computer}, 53(11):70--74, 2020.

\bibitem{Foidldatasmell}
Harald Foidl, Michael Felderer, and Rudolf Ramler.
\newblock Data smells: Categories, causes and consequences, and detection of
  suspicious data in ai-based systems.
\newblock {\em 2022 IEEE/ACM 1st International Conference on AI Engineering
  ^^e2^^80^^93 Software Engineering for AI (CAIN)}, pages 229--239, 2022.

\bibitem{frome2009large}
Andrea Frome, German Cheung, Ahmad Abdulkader, Marco Zennaro, Bo~Wu, Alessandro
  Bissacco, Hartwig Adam, Hartmut Neven, and Luc Vincent.
\newblock Large-scale privacy protection in google street view.
\newblock In {\em 2009 IEEE 12th international conference on computer vision},
  pages 2373--2380. IEEE, 2009.

\bibitem{guo2016ms}
Yandong Guo, Lei Zhang, Yuxiao Hu, Xiaodong He, and Jianfeng Gao.
\newblock Ms-celeb-1m: A dataset and benchmark for large-scale face
  recognition.
\newblock In {\em European conference on computer vision}, pages 87--102.
  Springer, 2016.

\bibitem{halevy2009unreasonable}
Alon Halevy, Peter Norvig, and Fernando Pereira.
\newblock The unreasonable effectiveness of data.
\newblock {\em IEEE intelligent systems}, 24(2):8--12, 2009.

\bibitem{hanley2020ethical}
Margot Hanley, Apoorv Khandelwal, Hadar Averbuch-Elor, Noah Snavely, and Helen
  Nissenbaum.
\newblock An ethical highlighter for people-centric dataset creation.
\newblock {\em arXiv preprint arXiv:2011.13583}, 2020.

\bibitem{Exposing.ai}
Jules. Harvey, Adam.~LaPlace.
\newblock Exposing.ai, 2021.

\bibitem{horne2000open}
Natasha~T Horne.
\newblock Open source software licensing: Using copyright law to encourage free
  use.
\newblock {\em Ga. St. UL Rev.}, 17:863, 2000.

\bibitem{2021towardsaccountability}
Ben Hutchinson, Andrew Smart, Alex Hanna, Emily Denton, Christina Greer, Oddur
  Kjartansson, Parker Barnes, and Margaret Mitchell.
\newblock Towards accountability for machine learning datasets: Practices from
  software engineering and infrastructure.
\newblock In {\em Proceedings of the 2021 ACM Conference on Fairness,
  Accountability, and Transparency}, pages 560--575, 2021.

\bibitem{kaminski2007open}
Halina Kaminski and Mark Perry.
\newblock Open source software licensing patterns.
\newblock {\em Computer Science Publications}, page~10, 2007.

\bibitem{pmlr-v80-kearns18a}
Michael Kearns, Seth Neel, Aaron Roth, and Zhiwei~Steven Wu.
\newblock Preventing fairness gerrymandering: Auditing and learning for
  subgroup fairness.
\newblock In Jennifer Dy and Andreas Krause, editors, {\em Proceedings of the
  35th International Conference on Machine Learning}, volume~80 of {\em
  Proceedings of Machine Learning Research}, pages 2564--2572, 10--15 Jul 2018.

\bibitem{kemelmacher2016megaface}
Ira Kemelmacher-Shlizerman, Steven~M Seitz, Daniel Miller, and Evan Brossard.
\newblock The megaface benchmark: 1 million faces for recognition at scale.
\newblock In {\em Proceedings of the IEEE conference on computer vision and
  pattern recognition}, pages 4873--4882, 2016.

\bibitem{khomh2018software}
Foutse Khomh, Bram Adams, Jinghui Cheng, Marios Fokaefs, and Giuliano Antoniol.
\newblock Software engineering for machine-learning applications: The road
  ahead.
\newblock {\em IEEE Software}, 35(5):81--84, 2018.

\bibitem{DataScientists}
Miryung Kim, Thomas Zimmermann, Robert DeLine, and Andrew Begel.
\newblock Data scientists in software teams: State of the art and challenges.
\newblock {\em IEEE Transactions on Software Engineering}, 44(11):1024--1038,
  2018.

\bibitem{li2021estimating}
Xiaoxiao Li, Ziteng Cui, Yifan Wu, Lin Gu, and Tatsuya Harada.
\newblock Estimating and improving fairness with adversarial learning.
\newblock {\em arXiv preprint arXiv:2103.04243}, 2021.

\bibitem{DataDebuggingFairness}
Yanhui Li, Linghan Meng, Lin Chen, Li~Yu, Di~Wu, Yuming Zhou, and Baowen Xu.
\newblock Training data debugging for the fairness of machine learning
  software.
\newblock {\em Proceedings of the 44th International Conference on Software
  Engineering}, page 2215^^e2^^80^^932227, 2022.

\bibitem{lo2021systematic}
Sin~Kit Lo, Qinghua Lu, Chen Wang, Hye-Young Paik, and Liming Zhu.
\newblock A systematic literature review on federated machine learning: From a
  software engineering perspective.
\newblock {\em ACM Computing Surveys (CSUR)}, 54(5):1--39, 2021.

\bibitem{MSdataset}
Madhumita Murgia.
\newblock {\em Microsoft quietly deletes largest public face recognition data
  set}.
\newblock [Last visited: 2022/11/27].

\bibitem{paullada2021data}
Amandalynne Paullada, Inioluwa~Deborah Raji, Emily~M Bender, Emily Denton, and
  Alex Hanna.
\newblock Data and its (dis) contents: A survey of dataset development and use
  in machine learning research.
\newblock {\em Patterns}, 2(11):100336, 2021.

\bibitem{pei2017deepxplore}
Kexin Pei, Yinzhi Cao, Junfeng Yang, and Suman Jana.
\newblock Deepxplore: Automated whitebox testing of deep learning systems.
\newblock In {\em proceedings of the 26th Symposium on Operating Systems
  Principles}, pages 1--18, 2017.

\bibitem{mitigating}
Kenny Peng, Arunesh Mathur, and Arvind Narayanan.
\newblock Mitigating dataset harms requires stewardship: Lessons from 1000
  papers.
\newblock {\em arXiv preprint arXiv:2108.02922}, 2021.

\bibitem{polyzotis2017data}
Neoklis Polyzotis, Sudip Roy, Steven~Euijong Whang, and Martin Zinkevich.
\newblock Data management challenges in production machine learning.
\newblock In {\em Proceedings of the 2017 ACM International Conference on
  Management of Data}, pages 1723--1726, 2017.

\bibitem{cani}
Gopi~Krishnan Rajbahadur, Erika Tuck, Li~Zi, Zhang Wei, Dayi Lin, Boyuan Chen,
  Zhen Ming, Daniel~Morales German, et~al.
\newblock Can i use this publicly available dataset to build commercial ai
  software? most likely not.
\newblock {\em arXiv preprint arXiv:2111.02374}, 2021.

\bibitem{rosen2005open}
Lawrence Rosen.
\newblock Open source licensing.
\newblock {\em Software Freedom and Intellectual Property Law}, 2005.

\bibitem{salay2017analysis}
Rick Salay, Rodrigo Queiroz, and Krzysztof Czarnecki.
\newblock An analysis of iso 26262: Using machine learning safely in automotive
  software.
\newblock {\em arXiv preprint arXiv:1709.02435}, 2017.

\bibitem{continuous}
Danilo Sato, Arif Wider, and Christoph Windheuser.
\newblock Continuous delivery for machine learning.
\newblock [Last visited: 2023/1/10].

\bibitem{dirtysecret}
Olivia Solon.
\newblock {\em Facial recognition’s ‘dirty little secret’: Millions of
  online photos scraped without consent}.
\newblock [Last visited: 2022/12/08].

\bibitem{stallman2002free}
Richard Stallman.
\newblock {\em Free software, free society: Selected essays of Richard M.
  Stallman}.
\newblock Lulu. com, 2002.

\bibitem{causingadecay}
Arjan Wijneveen.
\newblock {\em How copyright is causing a decay in public datasets}.
\newblock [Last visited: 2022/12/08].

\bibitem{yang2022study}
Kaiyu Yang, Jacqueline~H Yau, Li~Fei-Fei, Jia Deng, and Olga Russakovsky.
\newblock A study of face obfuscation in imagenet.
\newblock In {\em International Conference on Machine Learning}, pages
  25313--25330. PMLR, 2022.

\bibitem{zhang2010automatic}
Hongyu Zhang, Bei Shi, and Lu~Zhang.
\newblock Automatic checking of license compliance.
\newblock In {\em 2010 IEEE International Conference on Software Maintenance},
  pages 1--3. IEEE, 2010.

\bibitem{zhang2020machine}
Jie~M. Zhang, Mark Harman, Lei Ma, and Yang Liu.
\newblock Machine learning testing: Survey, landscapes and horizons.
\newblock {\em IEEE Transactions on Software Engineering}, 48(1):1--36, 2022.

\end{thebibliography}
\end{document}